\documentclass[fleqn,twoside]{article}
\usepackage{espcrc2}

\input epsf.sty
\newdimen\psfigsize

\def\etal{{\it et al.}}

\def\LP{\left(}		
\def\RP{\right)}	

\def\BE{\begin{equation}}
\def\EE{\end{equation}}
\def\BEA{\begin{eqnarray}}
\def\BEAN{\begin{eqnarray*}}
\def\EEA{\end{eqnarray}}
\def\EEAN{\end{eqnarray*}}

\newcommand{\ttred}{}
\newcommand{\ttblue}{}
\newcommand{\ttgreen}{}
 
\title{Spectrum Results with Kogut-Susskind Quarks}

\author{ Doug~Toussaint
\address{Department of Physics, University of Arizona, Tucson, AZ 85721, USA}
}

\begin{document}
\begin{abstract}
I summarize recent developments in spectrum calculations using
Kogut-Susskind quarks.  Theoretical developments include one-loop
computations with improved actions.  I present some recent simulation
results, mostly from a MILC collaboration project using three flavors.
Effects of dynamical quarks are clearly seen in the isovector $0^{++}$
meson propagator and in the mass ratio ``J''.
\end{abstract}
\maketitle

\renewcommand{\thefootnote}{\fnsymbol{footnote}}

The Kogut-Susskind formulation of lattice fermions, often called
``staggered fermions'', has long been a popular formulation for
QCD simulations including dynamical quarks.  The basic reason for
this is the single remaining exact chiral symmetry, which is
sufficient to protect the quark mass from additive renormalization.
From a practical viewpoint, an important consequence is a lower 
limit on the eigenvalues of $M^\dagger M$, which insures that the
simulation will be well behaved and not encounter ``exceptional
configurations.''  Also, but less importantly, there are simply
fewer fermionic degrees of freedom to handle.  The combined
effect of these advantages is that we can use our limited computer
power to push to smaller dynamical quark masses, or larger
physical sizes, or perhaps to larger statistics, by using
Kogut-Susskind quarks.  The price that we pay is that it can be
a painful exercise to figure out the lattice implementation of your
desired operator and, more importantly, flavor symmetry is
broken by effects of order $a^2$.

Although they are of order $a^2$, in practice the effects of
flavor symmetry breaking are unpleasantly large at accessible lattice
spacings.  The situation can be greatly improved by the use of
an improved action which suppresses the coupling of the quarks to
high momentum gluons, as well as fixing up the quarks' dispersion
relation. See Ref.~\cite{IMP_ACTION} for some relevant references.
The most important point is that exchange of a gluon with momentum
near $\pi/a$ can scatter a low momentum quark from one corner of
the Brillioun zone into another, resulting in a mixing of different
flavors.  ``Fattening'' the links, by averaging paths in the
parallel transport, effectively puts a form factor into the quark-gluon
vertex which suppresses such exchanges.


\begin{figure}[t]
\rule{0.0in}{0.3in}\vspace*{-0.3in}\\
\epsfxsize=3.00in
\epsfbox[0 0 4096 4096]{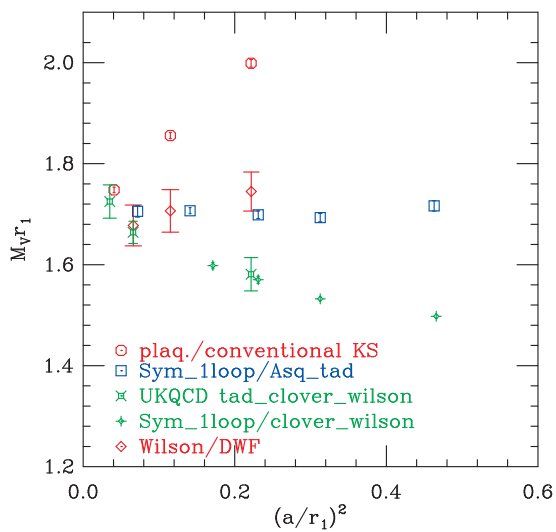} 
\rule{0.0in}{0.01in}\vspace{-0.5in}\\
\caption{
Mass of the quenched $\rho$ in units of $r_1$ at
$m_\pi r_1=0.807$.
The \ttblue squares are the improved Kogut-Susskind action\cite{MILCIMPRHO} and \ttred octagons the
conventional Kogut-Susskind action\cite{MILCCONVRHO}.  The \ttgreen plusses and fancy crosses
are clover-Wilson action\cite{WILSONRHO} and \ttred diamonds are domain wall quarks\cite{COLUMBIARHO}.
\label{SCALING_FIG}
}
\end{figure}


\begin{figure}[t]
\rule{0.0in}{0.3in}\vspace*{-0.3in}\\
\epsfxsize=3.30in
\epsfbox[0 0 4096 4096]{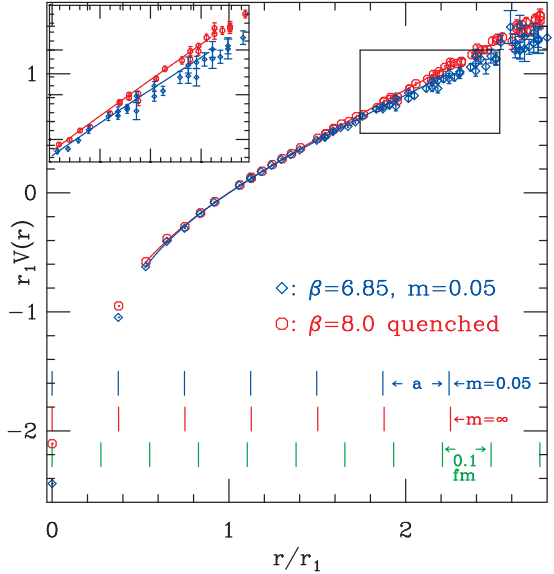} 
\rule{0.0in}{0.01in}\vspace{-0.5in}\\
\caption{
The static quark potential with $a \approx 0.13$ fm.
The \ttred octagons are the quenched potential and \ttblue diamonds the three
flavor potential at $m_s$.  The lines are fits to ``Coulomb
plus linear plus constant'', and the rulers show the lattice
spacings and units of 0.1 fm.
\label{POTMATCH_FIG}
}
\end{figure}

\begin{figure}[t]
\rule{0.0in}{0.3in}\vspace*{-0.3in}\\
\epsfxsize=2.90in
\epsfbox[0 0 4096 4096]{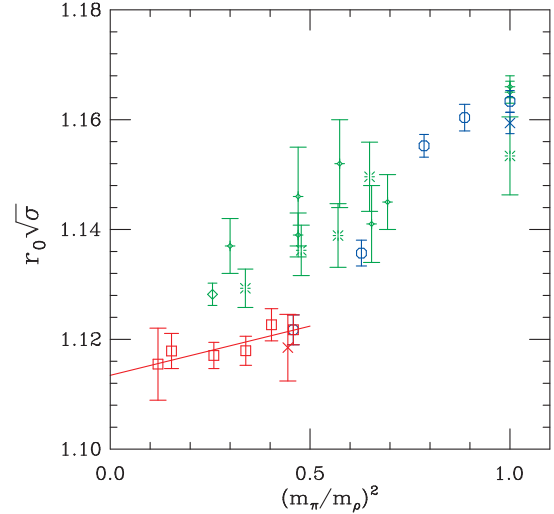} 
\rule{0.0in}{0.01in}\vspace{-0.5in}\\
\caption{
$r_0 \sqrt{\sigma}$ as a function of quark mass.  The quenched limit is
at the right side.  The \ttblue octagons are MILC results with three degenerate flavors
at $a \approx 0.13$ fm, \ttred squares are two light quarks and one strange
quark at $a \approx 0.13$ fm, and the \ttgreen diamond is a two flavor run
at 0.13 fm\protect\cite{MILC_POTENTIAL}.  The crosses are preliminary results at $a \approx 0.09$ fm.
The \ttgreen bursts are CPPACS results with two flavors of Wilson quarks\protect\cite{CPPACS_POTENTIAL}
and \ttgreen fancy diamonds are SESAM two-flavor Wilson results\protect\cite{SESAM_POTENTIAL}.
Some very new two-flavor Wilson results on smaller lattice (not plotted here)
which show a much smaller effect can be found in Ref.~\protect\cite{UKQCD_MATCHED}.
\label{R0_FIG}
}
\end{figure}


\begin{figure}[t]
\rule{0.0in}{0.3in}\vspace*{-0.3in}\\
\epsfxsize=3.00in
\epsfbox[0 0 4096 4096]{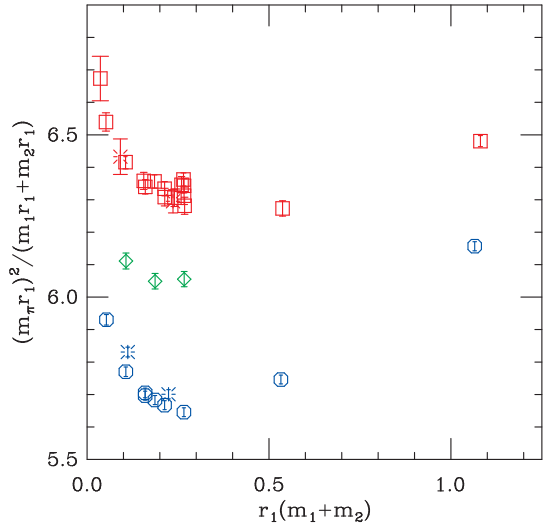} 
\rule{0.0in}{0.01in}\vspace{-0.6in}\\
\caption{
Squared pion mass divided by quark mass.
The \ttblue octagons are quenched $a \approx 0.13$ fm results,
\ttgreen diamonds two flavor, and \ttred squares three flavor.
The bursts are preliminary quenched and three-flavor
results at $a \approx 0.09$ fm.
\label{MPISQ_OVER_FIG}
}
\end{figure}

\begin{figure}[t]
\rule{0.0in}{0.3in}\vspace*{-0.3in}\\
\epsfxsize=3.00in
\epsfbox[0 0 4096 4096]{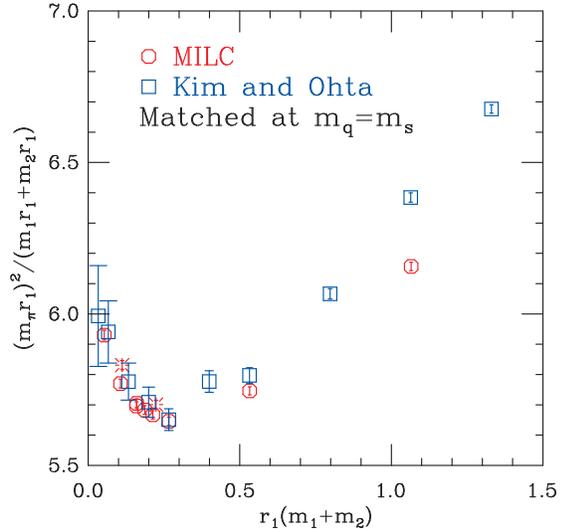} 
\rule{0.0in}{0.01in}\vspace{-0.6in}\\
\caption{
Quenched Goldstone pion masses from Kim and Ohta\protect\cite{KIM_AND_OHTA}
and MILC\protect\cite{MILC_IMP_SPECTRUM}.  The axes have been arbitrarily
rescaled to match the values at the strange quark mass, which
is approximately the low point on the curves.
\label{MPISQ_KO_FIG}
}
\end{figure}



\begin{figure}[t]
\rule{0.0in}{0.3in}\vspace*{-0.3in}\\
\epsfxsize=3.10in
\epsfbox[0 0 4096 4096]{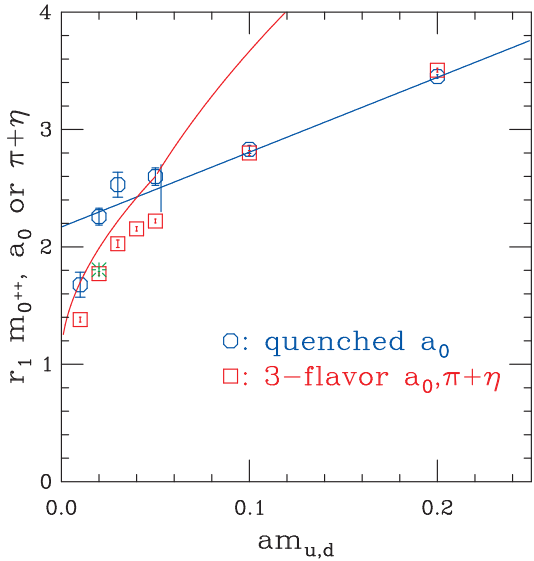} 
\rule{0.0in}{0.01in}\vspace{-0.6in}\\
\caption{
Masses for $0^{++}$ propagators in quenched QCD (\ttblue octagons)
and full QCD (\ttred squares).  The straight line is a linear fit
to the large mass points, intended to represent the mass of
a $q \bar q$ state, and the curved line is the mass of
a two particle $\pi+\eta$ state.
\label{A0_DECAY_FIG}
}
\end{figure}

At this conference we have seen the first results of one-loop calculations
done with improved Kogut-Susskind actions.
In Ref.~\cite{HEIN_TALK} the quark mass renormalization is calculated,
and in Ref.~\cite{TROTTIER_POSTER} renormalization constants for several
operators are shown.  The results are encouraging, with none of the
coefficients coming out surprisingly large.  In contrast, the conventional
Kogut-Susskind action often has large one-loop corrections.  This
can again be understood in terms of a form factor suppressing coupling
to high momentum gluons\cite{GOLTERMAN,HEIN_TALK}, so that a diagram like Fig.~2b
of Ref.~\cite{HEIN_TALK}
will not get large contributions when the quark is in another corner of
the Brillioun zone.  In other words, we suppress the unwanted doublers in
the loop diagrams.

Figure ~\ref{SCALING_FIG} shows the result of one scaling test of the
improved Kogut-Susskind action.  In this figure we plot the $\rho$ mass
in units of the static quark potential as a function of lattice spacing.
For the length scale we use $r_1$, the distance where $r_1^2 F(r_1)=1.0$.
All these points are from quenched simulations, and they are all interpolated
to the quark mass where $m_\pi r_1 = 0.807$, or $m_\pi \approx 460$ MeV.

Next one is motivated to ask whether it is practical to further improve
the Kogut-Susskind action.  At this conference DiPierro and Mackenzie
reported on experiments with empirically tuning the coefficients of the
paths in the action\cite{DIPIERRO_LAT}.  The gains were limited, which
they interpret as evidence of the need for four quark operators.
In Ref.~\cite{TROTTIER_POSTER} a one-loop improved action was presented,
and methods for handling these four quark operators by introducing auxiliary
bosonic fields were sketched.
Taking a different approach, A. Hasenfratz and Knechtli combined
fattening of the links with a projection back onto unitary SU(3)
matrices in the ``HYP'', for ``Hypercubic blocking'' action\cite{BOULDER_HYP}.  This
action produces smaller mass splittings among the pions than the
$a^2$ tadpole action.  These authors also describe an algorithm for
dynamical simulation of this action, which cannot be expressed in a
simple way as a sum over paths.

In another theoretical development presented at this conference,
Levkova and Manke have worked out an action for  unimproved dynamical Kogut-Susskind
quarks on an anisotropic lattice\cite{ANISOTROPIC_KS}.
While their primary motivation is high temperature QCD, this approach
could be useful for spectroscopy of glueballs, hybrids and
excited states.

The MILC collaboration has used the $a^2$ tadpole improved action for
a set of hadron spectrum calculations\cite{MILC_IMP_SPECTRUM}. These calculations used three
flavors of dynamical quarks, as well as a quenched run and one two-flavor
run, on lattices tuned to match the lattice spacing at about 0.13 fermi.
For masses larger than the strange quark mass $m_s$ three degenerate
flavors were used, while for smaller masses one quark mass was held fixed
at $m_s$.   Quark masses down to $m_q = 0.14 m_s$, or $m_\pi/m_\rho = 0.35$,
were used.
The lattice size was $20^3\times 64$, or $L \approx 2.6$ fm.
In addition, some preliminary results at a finer lattice spacing of $a \approx 0.09$ fm
are available.

The big advantage of using lattices matched in lattice spacing and physical
size for different numbers and masses of quarks is that the effects of dynamical
quarks can be convincingly exposed.  The simplest quantity that shows this is
the static quark potential.  Figure~\ref{POTMATCH_FIG} shows the potential in quenched
QCD and in three-flavor QCD with all the quarks at the strange quark mass.  The slopes of these two
potentials agree at the point chosen to define the length scale, but their overall
shape is different.  Notice that we do not see, and do not expect to see, string
breaking in the distance range shown here.  To quantify the change in shape we look
at dimensionless quantities such as $r_0 \sqrt{\sigma}$ or $r_0/r_1$.  Figure~\ref{R0_FIG}
shows $r_0 \sqrt{\sigma}$ for the three flavor runs.  It also contains two flavor
results with Wilson quarks from CPPACS and SESAM, and a two-flavor improved Kogut-Susskind
point from MILC.

One aspect of lattice spectroscopy where sea quarks are expected to
have important effects is in the chiral behavior of hadron masses.
In the case of the Goldstone pion, at lowest order the squared pion
mass is proportional to the quark mass, so corrections to this
behavior can be displayed by plotting the squared pion mass divided
by the quark mass.  Figure~\ref{MPISQ_OVER_FIG} is such a plot for the MILC
runs in both full and quenched QCD.  The increase at the right side
is understood as the transition from chiral behavior to heavy quark behavior,
while the much sharper upturn at low quark mass is the expected
chiral logarithm.  Although the quenched and three flavor curves look
similar, the theoretical expectation is that the three flavor line
has a finite limit as $m_q \rightarrow 0$, while the quenched line
diverges logarithmically.  The quenched pion masses in Fig.~\ref{MPISQ_OVER_FIG}
are similar to earlier quenched results of Kim and Ohta using the conventional
Kogut-Susskind action at a lattice spacing of about 0.46 fm\cite{KIM_AND_OHTA}.  A crude
comparison is in Fig.~\ref{MPISQ_KO_FIG}, where I have arbitrarily rescaled both
the vertical and horizontal axes to match the results at the strange
quark mass.

%
While it is tempting to simply fit Fig.~\ref{MPISQ_OVER_FIG} to the
chiral behavior predicted in Ref.~\cite{LEUTWYLER}, it is probably
necessary to take account of the remaining flavor symmetry
violations.
Claude Bernard has been fitting this data both directly to the
continuum form, and to a form that takes into account flavor
symmetry violations using the Lagrangian of Lee and Sharpe\cite{LEE_AND_SHARPE}
with empirically determined non-Goldstone pion masses in each loop,
to calculate corrections to the pion mass\cite{BERNARD_CHIRAL_FITS}.
In the quenched case, where the coefficient of the chiral logarithm
can be a parameter, reasonable fits can be obtained both with and
without the flavor symmetry breaking corrections.  However,
the coefficient of the logarithm comes out about 0.06 with the
simple fitting, and about 0.14 when the flavor symmetry breaking
is included.
When the three flavor data is fit, with the coefficients of the 
logarithms determined by the chiral theory, good fits can only be
obtained when the flavor symmetry breaking is included.

An application of these chiral corrections to the pion mass was
presented at this conference by the Ohio State group\cite{NELSON_LAT}.
Here a particular combination of chiral lagrangian parameters
was computed and found to be inconsistent with $m_{up}=0$.
The Boulder ``HYP'' action\cite{BOULDER_HYP}
was used for the valence quarks in this project.
In view of Bernard's observations on the effects of flavor
symmetry breaking, it is appropriate that they used a valence
quark action which makes these effects as small as possible.

The MILC spectrum calculations on matched quenched and three flavor
lattices\cite{MILC_IMP_SPECTRUM} found some interesting differences
between the quenched and full QCD spectra (and at least one interesting
lack of differences).  The most striking difference is in the coupling
of hadrons to two particle intermediate states, which is presumably 
present in full QCD calculations but is represented only by
``hairpin'' diagrams in quenched calculations.  
Figure~\ref{A0_DECAY_FIG} shows masses obtained for the isovector
$0^{++}$ propagator on matched quenched and full QCD lattices.
We interpret the low quark mass behavior of the full QCD results as
an avoided level crossing between a $q \bar q$ state and a two
meson $\pi+\eta$ state.  In simple language, we are seeing a
meson decay in the lattice simulations.

\begin{figure}[t]
\rule{0.0in}{0.3in}\vspace*{-0.3in}\\
\epsfxsize=3.00in
\epsfbox[0 0 4096 4096]{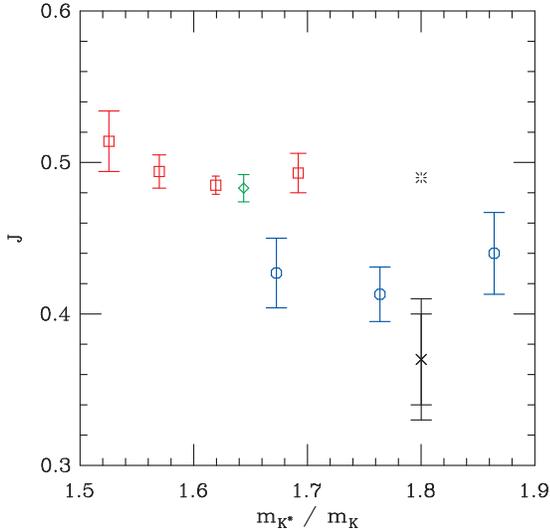} 
\rule{0.0in}{0.01in}\vspace{-0.6in}\\
\caption{
``J'' with improved Kogut-Susskind quarks.  The \ttblue octagons are quenched
values, \ttred squares three flavor and the \ttgreen diamond a two flavor point.
The burst is the real world value, and the cross the UKQCD
quenched value\protect\cite{UKQCD_J}.
The horizontal scale parameterizes the quark mass, with lighter
quarks to the right.
\label{J_FIG}
}
\end{figure}



\begin{figure}[t]
\rule{0.0in}{0.3in}\vspace*{-0.3in}\\
\epsfxsize=3.00in
\epsfbox[0 0 4096 4096]{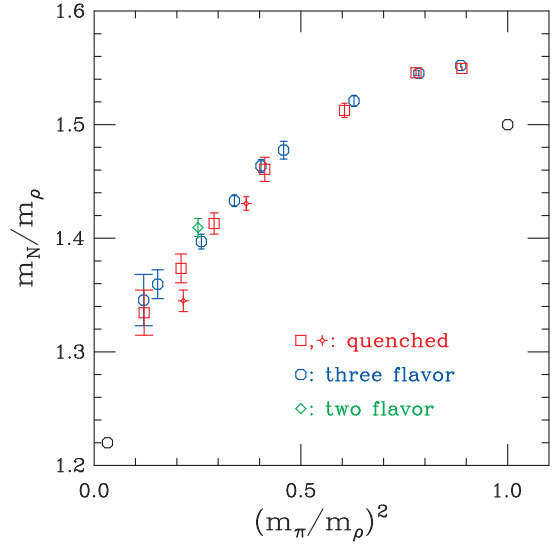} 
\rule{0.0in}{0.01in}\vspace{-0.6in}\\
\caption{
``Edinburgh plot'' for MILC matched quenched and full spectrum
calculations at $a \approx 0.13$ fm.  The \ttred squares are quenched, \ttblue octagons three flavor,
and the \ttgreen diamond two flavor results.
The \ttred plusses are quenched points at $a \approx 0.09$ fm.
\label{APE_BOTH_FIG}
}
\end{figure}

The UKQCD collaboration pointed out that the ratio ``J'', defined as
\BE J = m_{K^*} \frac{\partial m_V}{\partial m_{PS}^2}
\approx  \frac{m_{K^*}  \LP m_\phi-m_\rho \RP}{2 \LP m_K^2-m_\pi^2 \RP}
\EE
is a case where the quenched approximation produces an 
answer different from the real world\cite{UKQCD_J}.
Figure~\ref{J_FIG} shows this ratio on the matched quenched and
full QCD MILC lattices, with a clear increase in the value
when the sea quarks are included.  This result can be compared
with two flavor Wilson quark simulations by the CPPACS, UKQCD
and JLQCD collaborations\cite{CPPACS_POTENTIAL,UKQCD_MATCHED,JLQCD_J},
which also show a larger value in two flavor QCD.

The nucleon to rho mass ratio is a much studied quantity where lattice
simulations typically disagree with the real world number.
Many groups have looked at this, and it has become clear that
this quantity is especially sensitive to effects of the lattice
spacing.  Thus it is interesting to use the matched lattices to
look for effects of sea quarks.  Figure~\ref{APE_BOTH_FIG} shows
this ratio in the MILC calculations, with no discernable difference
between the quenched and full QCD curves.  This is in contrast
to UKQCD calculations on matched lattices with two flavors of
Wilson quarks, where the two flavor numbers are larger than the
quenched (See Fig.~9 of Ref.~\cite{UKQCD_MATCHED}).
The two plusses in Fig.~\ref{APE_BOTH_FIG} are quenched points
at $a \approx 0.09$ fm. Preliminary three-flavor results at
$a \approx 0.20$ and $0.09$ fm show a similar trend.  Although
this lattice spacing dependence is much less than seen with the
conventional Kogut-Susskind action, it is clear that a careful
continuum extrapolation will still be required for this quantity.
\rule{0.0in}{0.0in}\vspace*{-0.3in}\\

\section*{Acknowledgements}
I am grateful to the organizers of Lattice-01 for the
opportunity to present this talk.
I thank Claude Bernard, Joachim Hein, Takashi Kaneko, Francesco Knechtli, Kostas Orginos,
and Howard Trottier
for providing results used in this talk.
This work was supported by the U.S. DOE.

\end{document}